# Potassium carbonate under pressure: common structural trend for alkaline carbonates and binary compounds

Pavel N.Gavryushkin [1,2,*], Anna Y. Likhacheva [1], Zakhar I. Popov [3], Vladimir V. Bakakin [4], Konstantin D. Litasov [1,2], Anton F. Shatskiy [1,2], Alexey I. Ancharov [5] and Alex Gavryushkin [6]

[1] V.S. Sobolev Institute of Geology and Mineralogy, Russian Academy of Science, Siberian Branch, Novosibirsk, Russia

[2] Novosibirsk State University, Novosibirsk, Russia

[3] Kirensky Institute of Physics, Russian Academy of Science, Siberian Branch, Russia

[4] Nikolaev Institute of Inorganic Chemistry, Russian Academy of Science, Siberian Branch, Novosibirsk, Russia

[5] Budker Institute of Nuclear Physics, Russian Academy of Science, Siberian Branch, Novosibirsk, Russia

[6] Department of Computer Science, The University of Auckland, New Zealand

*Corresponding author, p.gavryushkin@gmail.com

**Abstract**

The behaviour of alkaline carbonates at high pressure is poorly understood. Indeed, theoretical and experimental investigations of general trends of pressure induced structural changes appear in the literature only sporadically. In this article we use a combination of *ab-initio* calculations and high-pressure experiments in diamond anvil cell to determine crystal structures of high-pressure phases of $K_2CO_3$. The comparison with experimental data on $Li_2CO_3$ allows to reconstruct the common structural trend, which is consistent with the simple rule that the structure of high-pressure polymorph is the same as the ambient structure of a heavier element compound from the same group of the periodic table. The correctness of the constructed trend is confirmed by the perfect consistency with pressure-induced structural transformations in a wide range of binary $A_2B$ compounds.

# Introduction

Potassium carbonate or potash is a well-known compound of great importance in the everyday life. While potash is not presented in the Earth's interior as a separate phase, it could play a key role in mantle melting, plume upwelling and diamond growth. For example, adding a small amount of this compound to mantle rocks decreases their melting temperature by several hundred degrees [1,2] and substantially decreases the energetic barrier of diamond growth [3].

The structural trend of high-temperature phases of all alkaline carbonates are investigated in detail and well understood [4,5,6]. However, almost nothing is known about high-pressure behavior of these compounds. We are aware of only one experimental work on $Li_2CO_3$ [7] and one theoretical work on crystal structure prediction for a range of alkali carbonates [8]. The experimental results show that the high pressure phase of $Li_2CO_3$ does not have structural analogues among heavier element alkaline carbonates. This means that high-pressure phases of alkaline carbonates do not have a common structural trend at all or this trend is more complex than that of alkaline earth carbonates and various other compounds, where high-pressure phase of the lighter element compound is isostructural to ambient phase of the heavier element compound. The theoretical investigation [8] mentioned above does not reveal any common features in high-pressure structures of alkaline carbonates, which can be considered as an additional argument against the existence of a common high-pressure trend.

In the present work, we employ crystal structure prediction evolutionary algorithms and in-situ X-ray experiments in diamond anvil cell (DAC) to provide clear evidences for the existence of a common trend for alkaline carbonates. Furthermore, we show that the trend is consistent with those of various $A_2B$ binary compounds.

There are two principal structural types of alkaline carbonates: $Li_2CO_3$ and $K_2CO_3$. Other carbonates—$Rb_2CO_3$ and $Cs_2CO_3$—are isostructural to $K_2CO_3$ whereas $Na_2CO_3$ is different only in the tilt of $CO_3$ triangles. Since the high-pressure phase is known for $Li_2CO_3$, to understand the general structural trend we have to determine the high-pressure phases of one of the compounds of other structural type. We choose potassium carbonate as a representative of such compounds..

# Results

**Ab-initio calculations**

To test the methodology we employ, the ambient γ-$K_2CO_3$ structure was predicted. We report that the atomic coordinates and unit cell parameters are in good agreement with the experimental values [9].

Three new phases were predicted under high-pressure: $P2_1/c$, $C2/c$ и P-1. All these phases are structural relatives to the high-pressure polymorph of $Li_2CO_3$ (Fig. 1). Their unit cell parameters and atomic coordinates are presented in the *Supplementary materials*. Phonon dispersion curves indicate that these structures are dynamically stable.

The pressure dependence of energies for predicted phases are shown in Fig.2. According to our calculations, the sequence of phase transitions is the following: 12 GPa, γ-$K_2CO_3$ → P-1; 53.5 GPa P-1→ $C2/c$. Among other predicted structures, three structures have analogs among alkaline carbonates: γ-$K_2CO_3$, β-$Na_2CO_3$-type and $Li_2CO_3$-HP-type. However, only γ-$K_2CO_3$ has a region of thermodynamic stability (Fig. 2).

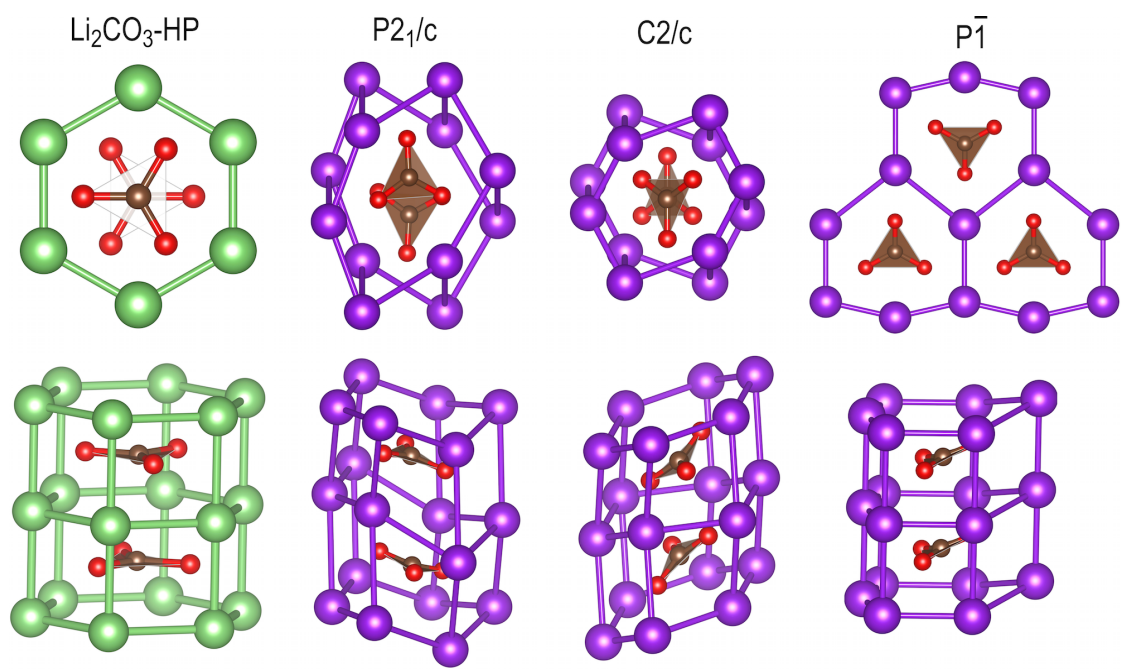

**Fig. 1.** High-pressure structures of $Li_2CO_3$ and $K_2CO_3$. Structures in upper and lower rows are shown in different orientations. VESTA software [10] was used for structure visualization.

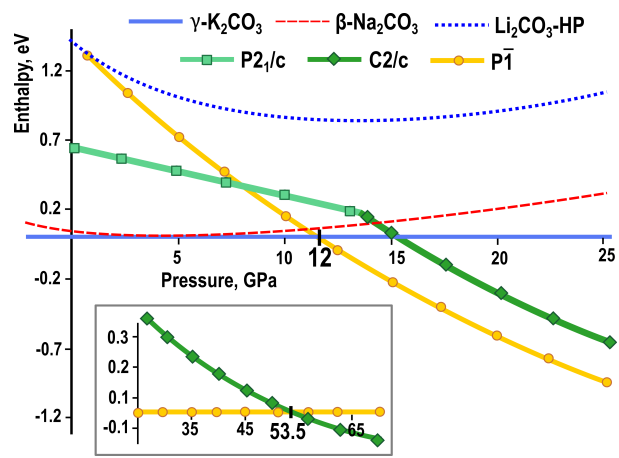

**Fig. 2.** Phase diagram of $K_2CO_3$ from *ab-initio* calculations.

**Experiment**

The lack of experimental data on potassium and other alkaline carbonates can partially be explained by their high hygroscopicity. The hygroscopicity of potassium carbonate is so high that the carbonate almost immediately transforms to sesquihydrate $K_2CO_3*1.5H_2O$ at contact with air [11]. Peaks of sesquihydrate are overlapped with peaks of $\gamma$-$K_2CO_3$ and greatly obstruct the structure solution. To overcome this difficulty we used powder diffraction methods in conjunctions with ab-initio DFT calculations and crystal structure prediction.

In the pressure range 2.3-3.1 GPa the diffraction pattern undergoes substantial changes, which indicates the formation of the high-pressure phase. This phase becomes amorphous upon decompression to the ambient pressure (Fig.3).

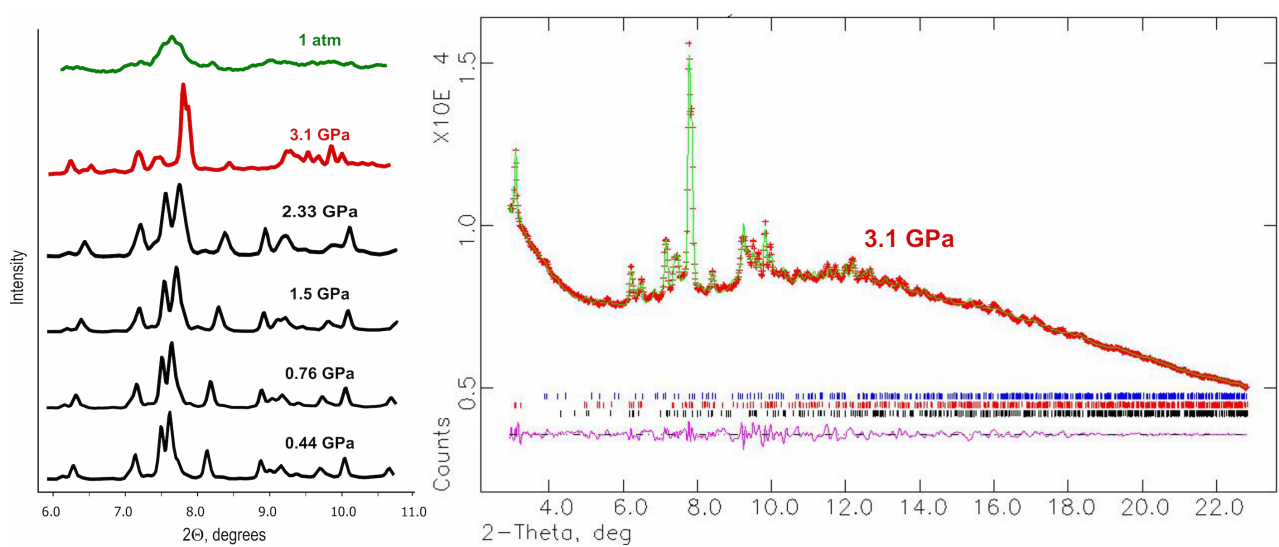

**Fig. 3.** High-pressure diffraction patterns of potassium carbonate. Left figure - consecutive diffraction patterns of $K_2CO_3$ under compression and decompression (top graph). Right figure – results of Rietveld refinement of theoretical $P2_1/c$-structure; experimental and calculated diffraction profiles are indicated by '+' and by continuous line, respectively. Tick marks indicate positions of allowed Bragg peaks for $P2_1/c$-structure (blue), $K_2CO_3*1.5H_2O$ (red) and $\gamma$-$K_2CO_3$ (black). Difference curve is plotted on the bottom graph.

The comparison of the diffraction pattern of high pressure phase with model diffraction patterns of phases predicted by [8] shows that the experimental pattern does not match any of them. Among structures founded in our calculations, the $P2_1/c$-structure shows relevant matching with the experimental diffraction pattern at 3.1 GPa. The results of Rietveld refinement of this structure against experimental data are shown in Fig. 3. The refinement reveals that in addition to the high-pressure $P2_1/c$-phase, minor amounts (less than 5%) of sesquihydrate and γ-$K_2CO_3$ present in the sample as well. The final R-factors were calculated to be: GOF=0.99, $R_p$=0.008, $R_F^2$=0.14. The unit cell parameters and atomic coordinates of $P2_1/c$-phase are presented in Table 1. The refinement leaves atomic coordinates almost the same, although it slightly changes the theoretical unit cell parameters.

## Discussion

In the case of potassium carbonate, classical description of crystal structures based on anion polyhedrons is complicated due to the presence of rigid $CO_3$ groups and complex low-symmetry cation arrangement. As a result, the approach based on anion polyhedrons does not allow to combine the high-pressure and low-pressure polymorphs into a general structural trend. The cation net approach [4, 12] is more fruitful in this case. In addition to making the structure description more easy and clear, this method also allows to find an analogy between structural changes of alkaline carbonates and binary $A_2B$ compounds.

All alkaline carbonates except for $Li_2CO_3$ have the same underlying net of $Ni_2In$-type. The underlying net of $Li_2CO_3$ is of anti-$CaF_2$-type and the underlying net of its high-pressure polymorph is of $AlB_2$-type [4].

The high-pressure structures of potassium carbonate found in our calculations are close to one another. *All structures we found can be described as deformed HP-$Li_2CO_3$-structures. All of them have an array of hexagonal prisms of potassium anions centered by $CO_3$ triangles and sharing common faces. The orientation of $CO_3$ triangles inside the prism can be different for different polymorphs but cation arrays are in principle the same. Different polymorphs—$P2_1/c$, $C2/c$ and $P-1$—are caused by different distortions of this array.* It is worth noting that the energy of non-deformed $Li_2CO_3$-HP structure – the archetype of the high-pressure polymorphs of $K_2CO_3$ – is substantially higher than the energy of the other polymorphs (Fig.2) in all pressure range. Therefore this polymorph should not be realized in the case of potassium carbonate.

Among three baric phases of $K_2CO_3$ only two are substantially different. $P2_1/c$- and $C2/c$-structures are connected by a continuous phase transition. The distance between the carbon atoms in the $P2_1/c$-polymorph is decreased during compression until the carbon atoms are not placed exactly one under another (Fig.1) as in the $C2/c$-structure.

Let us compare the sequences of phases of $Li_2CO_3$ and $K_2CO_3$ under compression:

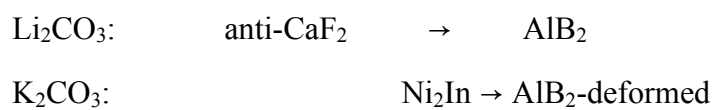

$Li_2CO_3$:　　　　anti-$CaF_2$　　→　　$AlB_2$

$K_2CO_3$:　　　　　　　　$Ni_2In$ → $AlB_2$-deformed

Both trends finish up in the same structural type, which is $AlB_2$. However, the $Ni_2In$-phase is missed out in the case of lithium carbonate, which complicates the determination of the common trend for alkaline carbonates based solely on $Li_2CO_3$ data. Indeed, we are aware of only one experimental work on $Li_2CO_3$ [7] which claims that the $Ni_2In$-structure is missed out from the sequence of phases. We expect however that this structure will be synthesized in future experiments.

Combining these two trends, for $Li_2CO_3$ and $K_2CO_3$, we suggest the full sequence of pressure induced structural changes as:

$$anti\text{-}CaF_2 \rightarrow Ni_2In \rightarrow AlB_2$$

*The first part of this trend corresponds to the well-known rule according to which the structure of a high-pressure polymorph is the same as the ambient structure of a heavier element compound in the same group of the periodic table.* Thus, the behavior of alkaline carbonates is the same as that of alkaline-earth carbonates.

The analysis of high-pressure experimental and theoretical data [13, 14, 15] indicates that this trend is reproduced on a wide range of alkaline compounds including binary sulfides, oxides, selenides, tellurides and related oxides (sulfates, selenates and tellurates)). Under compression anti-$CaF_2$-structures of these compounds transform into $Ni_2In$-structures, which transform into $AlB_2$-type structures according to crystal structure prediction calculations.

This remarkable correspondence of trends for such a great variety of different compounds provides new arguments for the possibility of synthesising of a $AlB_2$-type structure in the case of $A_2B$ compounds. Furthermore, the correspondence confirms the correctness of our experimental and theoretical results on alkaline carbonates providing a background for further high-pressure experiments.

## Methods

**Ab-initio calculations.** Crystal structure prediction has been performed using evolutionary algorithms implemented in USPEX (Universal Structure Predictor: Evolutionary Xtallography) package [16, 17, 18, 19, 20]. Local optimization has been performed in the framework of DFT theory with VASP code [21, 22], using the plane wave basis set and projector augmented wave method [23]. Exchange-correlation effects were taken into account in the generalized gradient approximation (GGA) using Perdew–Burke–Ernzerhof (PBE) functional [24]. For the crystal structure search we used a plane-wave basis set cutoff of 520 eV and performed the BZ integrations using uniform

Gamma-centred k-point meshes with a k-point grid of spacing $2\pi \times 0.025$ Å$^{-1}$. Iterative relaxation of atomic positions was stopped when all forces were smaller than 0.005 eV Å$^{-1}$. Calculations were carried out for 0, 30 and 60 GPa with 1, 2, 3, 4, 5 and 6 formula units per unit cell. The temperature in all calculations was 0 K.

**Structure refinement.** The unit cell parameters and atomic coordinates were refined by Rieveld method with GSAS software package [25] in isotropic approximation.

**Materials.** Symmetric DACs equipped with flat anvils were used to generate pressures up to 4 GPa. The cullet size is 300 μm. A sample of the synthetic reagent $K_2CO_3$ (99.99 %, , Wako Co Ltd) was ground in agate mortar and mixed with silicon oil in the ratio of 1:4. This mixture was loaded in a 400 × 100 μm sample cavity in a rhenium gasket of 40–50 mm thickness under ambient conditions and compressed to 0.3 GPa. Powder sample was preliminarily annealed in a vacuum oven at 473 K for 2 hours to minimize the formation of sesquihydrate.

**X-Ray diffraction measurements.** *In situ* X-ray diffraction experiments were conducted at the beamline #4 of the VEPP-3 storage ring of the SSRC Synchrotron Centre (Novosibirsk, Russia) [26]; X-ray wavelength was 0.3685 Å. MAR345 imaging plate detector (pixel dimension 100 mm) was used for data collecting. FIT2D [27] program was used to integrate the two-dimensional images to a maximum 2Θ value of 25º. The pressure was measured by displacement of $^5D_0$–$^7F_0$ fluorescence line of $SrB_4O_7$:$Sm^{2+}$ with PRL (BETSA) spectrometer with accuracy ~ 0.05 GPa. Experiments were carried out in the pressure range 0.45 – 3.1 GPa at 298 K.

**Author contributions.** PNG, AFS and KDL designed the study, AYL, PNG AIA carried out X-ray diffraction experiments, AYL performed the structure refinement, PNG, ZIP and AG performed calculations, PNG and VVB perform crystalchemical analysis of structures, PNG wrote the manuscript. All authors proofread the manuscript.

**Acknowledgments**

We thank the Information Technology Centre of Novosibirsk State University for providing access to the cluster computational resources and the SSRC Synchrotron Centre for a possibility of performing the experiments.

**Funding sources**

The research was supported by the Russian Foundation for Basic Research through grant (No 14-05-31051) and the Ministry of Education and Science of Russian Federation (No 14.B25.31.0032 and MK-3766.2015.15).

Table 1. Refined atomic coordinates and unit cell parameters of $P2_1/c$-structure.

Space group $P2_1/c$
a = 8.488(7) Å, b = 6.032(9) Å, c = 6.881(5) Å, β = 102.33(4)°

|    | x | y | z |
|----|---|---|---|
| K1 | 0.434(4) | 0.609(12) | 0.214(9) |
| K2 | 0.105(4) | 0.889(9) | 0.326(10) |
| C  | 0.7524(26) | 0.8531(30) | 0.997(11) |
| O1 | 0.3690(28) | 0.272(5) | 0.993(10) |
| O2 | 0.1165(26) | 0.247(6) | 0.018(10) |
| O3 | 0.745(6) | 0.0654(29) | 1.008(16) |